\documentclass[3p,times]{elsarticle}

\usepackage{amssymb}
\begin{document}

\begin{frontmatter}


%% \author{Hannu Holopainen\corref{cor1}\fnref{label2}}
%% \ead{hannu.l.holopainen@jyu.fi}


\title{Elliptic flow from event-by-event hydrodynamics with fluctuating initial state}


\author[add1,add2]{Hannu Holopainen (speaker)}
%\ead{hannu.l.holopainen@jyu.fi}
\author[add3]{Harri Niemi}
%\ead{niemi@th.physik.uni-frankfurt.de}
\author[add1,add2]{Kari J. Eskola}
%\ead{kari.eskola@phys.jyu.fi}

\address[add1]{Department of Physics, P.O.Box 35, FIN-40014 University of Jyv\"askyl\"a, Finland}
\address[add2]{Helsinki Institute of Physics, P.O.Box 64, FIN-00014 University of Helsinki, Finland}
\address[add3]{Frankfurt Institute for Advanced Studies, Ruth-Moufang-Str. 1, D-60438 Frankfurt am Main, Germany}

\begin{abstract}
We develop an event-by-event ideal hydrodynamical framework where initial state density
fluctuations are present and where we use a similar flow-analysis method as in the
experiments to make a one-to-one $v_2$ comparison with the measured data. Our studies
also show that the participant plane is quite a good approximation for the event plane. 
\end{abstract}


\end{frontmatter}

% \linenumbers

\section{Introduction}
\label{sec: intro}
Hydrodynamical calculations based on averaged initial conditions typically
have problems in reproducing the experimentally observed centrality- and
$p_T$-dependence of elliptic flow coefficient $v_2(p_T)$ at RHIC, see e.g.
Fig 7.5 in Ref.~\cite{Niemi:2008zz}. Especially, in the most central collisions
$v_2(p_T)$ is clearly underestimated. In this talk I will show that this problem
can be solved by using ideal event-by-event hydrodynamics \cite{Holopainen:2010gz}.
This model reproduces the measured $v_2(p_T)$ for all centrality classes from
0-5\% to 30-40\% up to $p_T \sim 2$ GeV. 

\section{Event-by-event hydrodynamics framework}
\label{sec: ebye hydro}

\begin{figure}[t]
  \begin{minipage}{8cm}
    \centering
    \vspace{-0.4cm}
    \includegraphics[height=7.0cm]{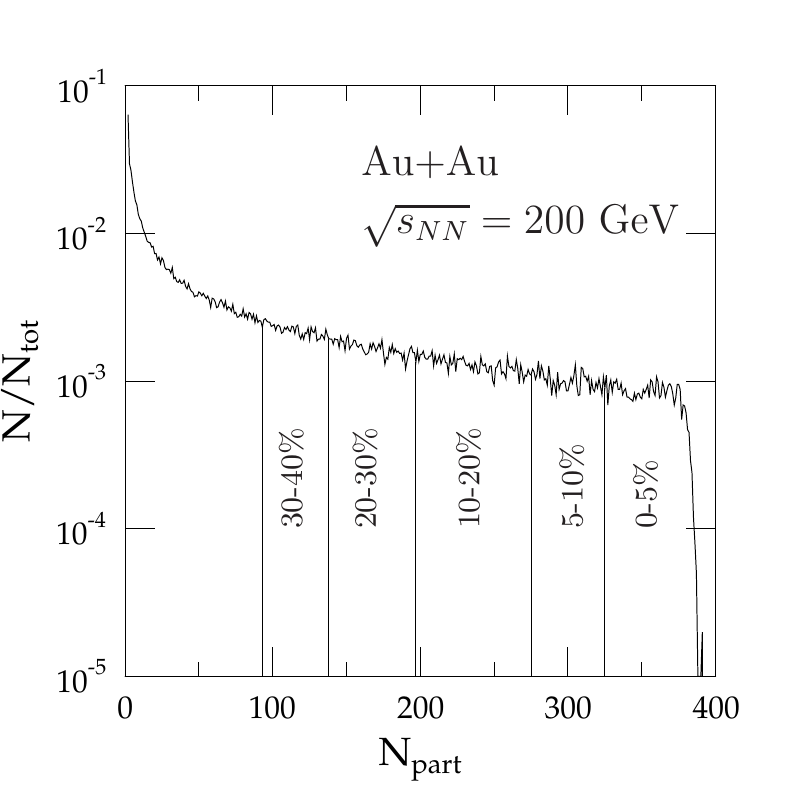}
    \caption{\protect\small Our centrality class definition for Au+Au
           collisions at $\sqrt{s_{NN}} = 200$~GeV in terms
           of the number of participants. From \cite{Holopainen:2010gz}.}
    \label{fig: centrality class}
  \end{minipage}
  \hfill
  \begin{minipage}{8cm}
    \centering
    \includegraphics[height=7.0cm]{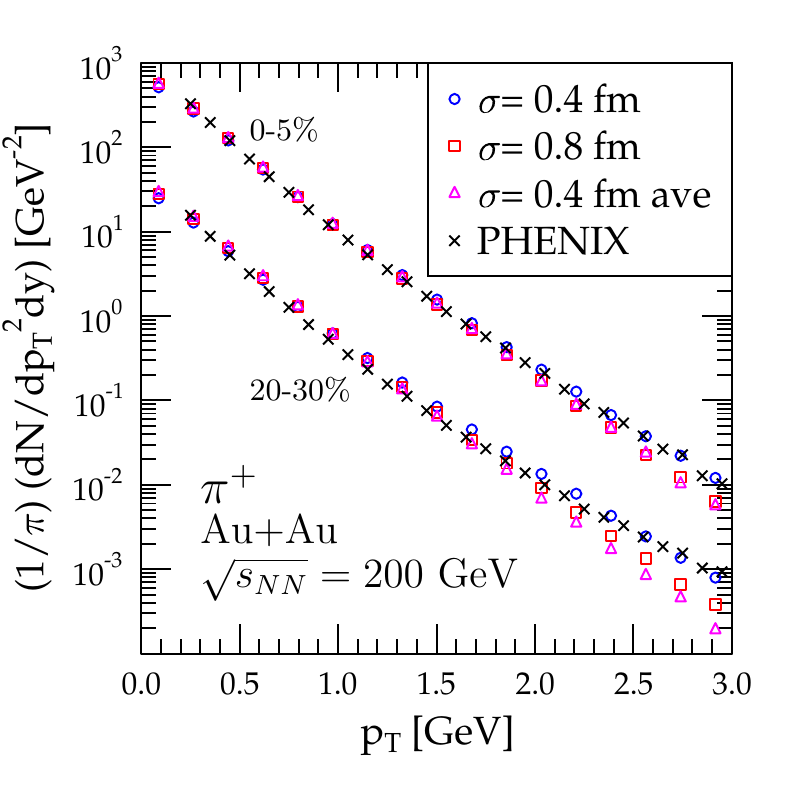}
    \caption{\protect\small Positive-pion $p_T$ spectrum for Au+Au
           collisions at $\sqrt{s_{NN}} = 200$ GeV calculated with averaged
           and fluctuating initial states varying the Gaussian smearing width $\sigma$.
           Data are from PHENIX \cite{Adler:2003cb}. From \cite{Holopainen:2010gz}.}
    \label{fig: pion spectra}
  \end{minipage}
\end{figure}

Initial states are generated here with a Monte Carlo Glauber (MCG) model. First we distribute
the nucleons into nuclei using the standard Woods-Saxon density distribution. In the transverse
plane the two colliding nuclei are separated by an impact parameter $b$ which is sampled
from a distribution $dN/db \propto b$. Nucleons $i$ and $j$ from different nuclei collide
if
\begin{equation}
  (x_i - x_j)^2 + (y_i - y_j)^2 \le \frac{\sigma_{NN}}{\pi},
\end{equation}
where $\sigma_{NN}$ is the
inelastic nucleon-nucleon cross section. For collisions at $\sqrt{s_{NN}} = 200$ GeV,
considered here, we take $\sigma_{NN} = 42$ mb. For simplicity we do not include any
nucleon finite size effects.
As shown in Fig.~\ref{fig: centrality class} in order to define centrality classes
we slice the distribution of $N_{\rm part}$ so that each interval contains a certain percentage
of total events. Thus the impact parameter may vary freely in each centrality class.

MCG gives only the positions of wounded nucleons, but we must start our hydro with
an energy density profile. We have chosen to distribute the energy density around wounded
nucleons using a 2D Gaussian as a smearing function,
\begin{equation}
  \epsilon (x,y) = \frac{K}{2\pi \sigma^2} \sum_{i=1}^{N_{\rm{part}}} \exp\Big( -\frac{(x-x_i)^2+(y-y_i)^2}{2\sigma^2} \Big),
  \label{eq:eps}
\end{equation}
where $\sigma$ is a free smearing parameter controlling the width of the Gaussian.
The overall normalization constant $K$ and the initial time
$\tau_0 = 0.17$ fm are taken from the EKRT pQCD+saturation model \cite{Eskola:1999fc}.

We solve the standard ideal hydrodynamic equations $\partial_\mu T^{\mu\nu} = 0$ numerically.
We have neglected the small net-baryon density since we are interested only in particle production
at mid-rapidity. For the same reason we can assume longitudinal boost-invariance, which
reduces the numerical problem to (2+1)-dimensions. To be able to solve the hydrodynamical
equations we need an Equation of State (EoS) to relate pressure and energy density. Our
choice is the EoS from Ref.~\cite{Laine:2006cp}.

We use the conventional Cooper-Frye method to calculate thermal spectra of hadrons
emitted from a constant-temperature surface $T_{\rm dec}=160$~MeV.
To get the individual final state particles for the flow analysis, we sample the
thermal spectrum. We must also take into account all strong and electromagnetic
decays before we can compare with the data. For this, we let the unstable thermal
hadrons decay one by one using PYTHIA 6.4 \cite{Sjostrand:2006za}.

Elliptic flow of hadrons is calculated here with respect to two different reference
planes. The simplest way is to use the reaction plane, defined by the impact parameter
and beam axis, as a reference plane. Another way to calculate $v_2$ is the event plane
method \cite{Poskanzer:1998yz}, where one determines an event flow vector for each event
\begin{equation}
  Q_2 = \sum_i ( p_{Ti} \cos(2\phi_i),  p_{Ti} \sin(2\phi_i) ),
\end{equation} 
where we sum over every particle in the event and where $\phi_i$ is measured from the
$x$ axis, which is here fixed by the impact parameter. The event plane angle $\psi_{2}$
for each event is then defined to be
\begin{equation}
  \psi_2 = \frac{ {\rm arctan} ( Q_{n,y} / Q_{n,x} ) }{2}.
\end{equation}
Then the ''observed'' elliptic flow, $v_2\{{\rm obs}\} = \langle \cos (2(\phi_i-\psi_2)) \rangle $, can be calculated with respect to the event plane. However,
the event plane fluctuates around the ''true'' event plane due to a finite number
of particles from which the event plane is determined. We apply the two-subevent method
\cite{Poskanzer:1998yz} to estimate the event plane resolution $\mathcal{R}_2$, which
corrects $v_2 \{ {\rm obs} \}$ for the event plane fluctuations. The
final event plane elliptic flow is obtained as
\begin{equation}
  v_2\{ {\rm EP} \} = v_2\{{\rm obs}\}/\mathcal{R}_2.
\end{equation}

\begin{figure}[th]
  \centering
  \includegraphics[height=13.5cm]{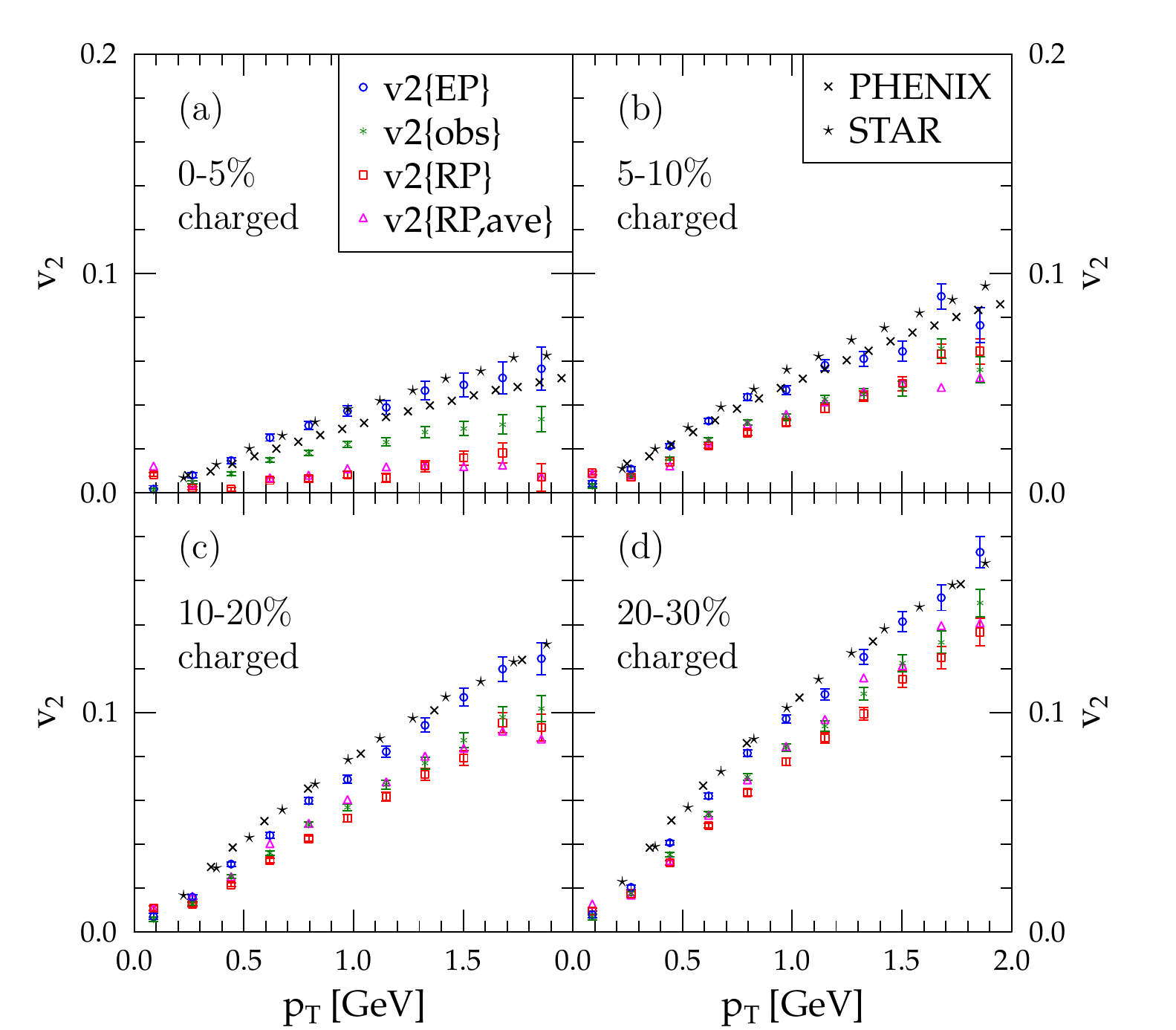}
  \caption{\protect\small Elliptic flow of charged particles as a function of $p_T$
           at different centralities in Au+Au collisions at $\sqrt{s_{NN}} =
           200$ GeV. Hydrodynamical results shown are for 
           $\sigma = 0.4$ fm. Data are
           from PHENIX \cite{Afanasiev:2009wq,Adare:2010ux} and STAR
           \cite{Adams:2004bi}. The statistical errors in the experimental data
           are smaller than the symbol size.}
  \label{fig: differential v2}
\end{figure}

\section{Results}
\label{sec: results}

We have plotted in Fig.~\ref{fig: pion spectra} the transverse momentum spectra of
positively charged pions from three hydro calculations. We can see that we get more
particles at high $p_T$ with fluctuating initial states and smaller $\sigma$,
in which case larger pressure gradients are present.
Most importantly, we reproduce the $p_T$ spectra sufficiently well,
so that we can meaningfully study the elliptic flow.

In Fig.~\ref{fig: differential v2} we plot the elliptic flow of charged particles
as a function of $p_T$ at four different centrality classes. We compare the
event-by-event calculations $v_2\{{\rm RP}\}$ and $v_2\{{\rm EP}\}$ and $v_2\{{\rm obs}\}$ with
$v_2\{{\rm RP,ave}\}$ which is calculated from an averaged initial state. We remind,
though, that $v_2\{{\rm obs}\}$ is a non-physical quantity since it
depends e.g. on the number of particles (rapidity acceptance) used in the calculation. 

We see that $v_2\{{\rm RP}\}$ and $v_2\{{\rm RP,ave}\}$ are very similar at all centralities. This
means that fluctuations alone do not generate more elliptic flow. Especially this means
that the $v_2$ deficit persists in most central collisions even with fluctuations.
However, we see that $v_2\{{\rm EP}\}$ fits the measured data surprisingly well at all
centralities shown here up to $p_T \sim 2$ GeV. Thus we can conclude that the
reference plane definition is indeed very important in $v_2$ calculations and thus
also in viscosity determination from the measured $v_2$ data.

\begin{figure}[t]
  \begin{minipage}{8cm}
    \centering
    \includegraphics[height=7.0cm]{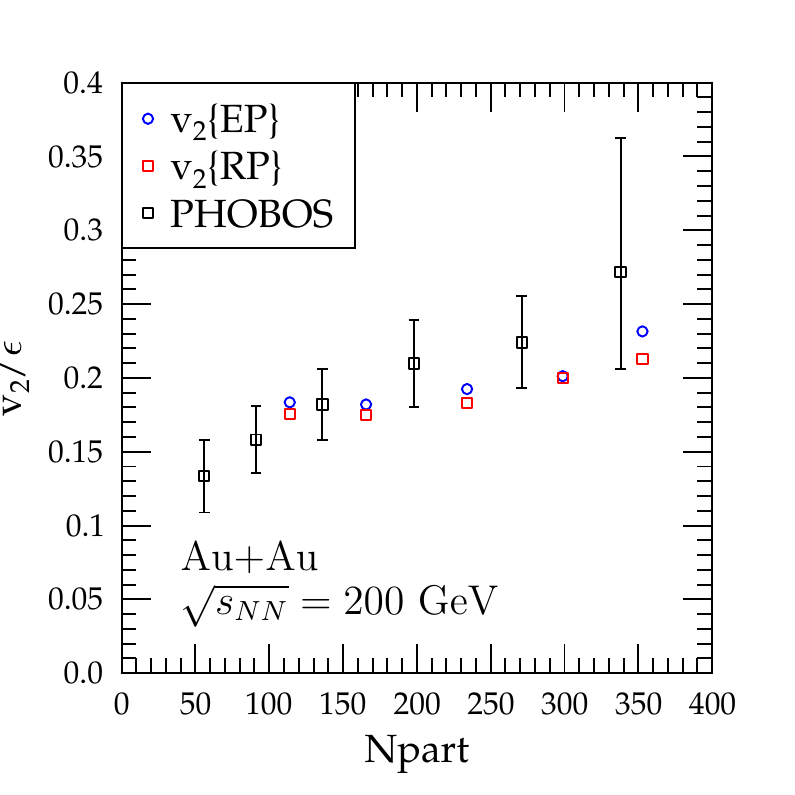}
    \caption{\protect\small Integrated elliptic flow of charged particles
           in Au+Au collisions at $\sqrt{s_{NN}} = 200$~GeV divided by initial
           eccentricity. The data from PHOBOS
           \cite{Alver:2006wh}. From \cite{Holopainen:2010gz}.}
    \label{fig: int v2/ecc}
  \end{minipage}
  \hfill
  \begin{minipage}{8cm}
    \centering
    \includegraphics[height=7.0cm]{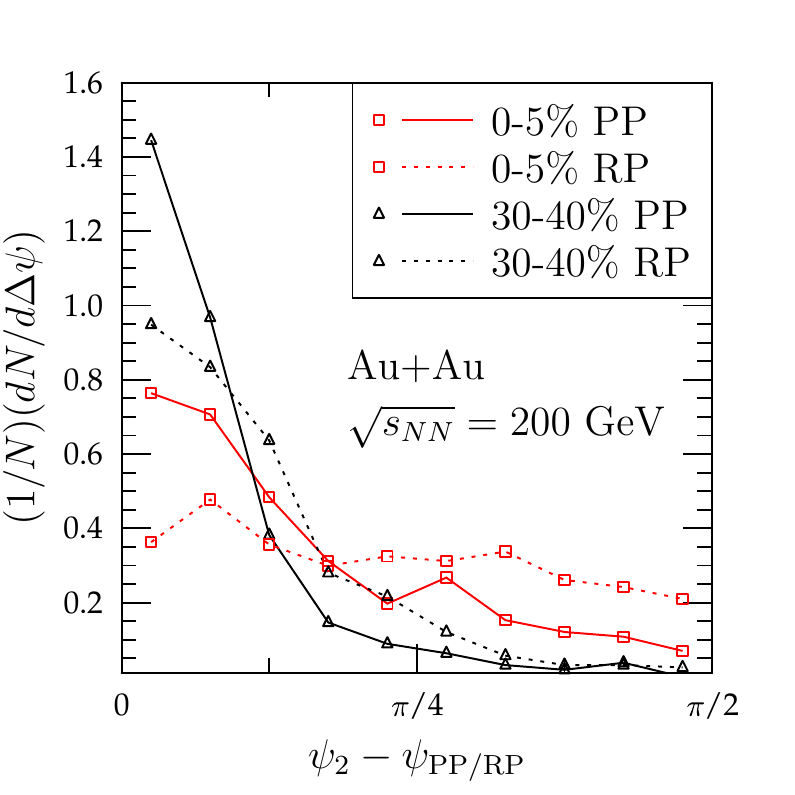}
    \caption{\protect\small Correlation of the event plane with the participant plane, and with the reaction plane at two different centralities. The lines are to guide the eye.
    From \cite{Holopainen:2010gz}.}
    \label{fig: ep vs rp}
  \end{minipage}
\end{figure}

In Fig.~\ref{fig: int v2/ecc} we have plotted the elliptic flow divided by the
initial eccentricity. In the data and in our $v_2\{{\rm EP}\}$ points, $v_2\{{\rm EP}\}$ is divided
by the participant eccentricity, which is the maximal eccentricity of initial matter
distribution. We have also plotted $v_2\{{\rm RP}\}$ divided by the eccentricity calculated
with respect to the reaction plane. We notice that both calculations are in an
agreement with the data and each other.

We have also studied how well the event plane and participant plane are correlated.
In Fig.~\ref{fig: ep vs rp} we have plotted the distribution of events as a function of
the angle difference from event plane to the participant plane and to the reaction plane.
We see that the event plane is indeed better correlated with the participant 
plane than with the reaction plane. This suggest, as intuitively expected, that the participant plane is quite a
good approximation for the event plane.

\section*{Acknowledgments}
HH and KJE thank the Vilho, Yrj\"o and Kalle V\"ais\"al\"a fund of the Finnish Academy of Science and Letters, and the Academy of Finland (Project nr. 133005) for financial support.



%% References


\bibliographystyle{elsarticle-num}
\bibliography{proc}






% \end{thebibliography}

\end{document}